# Luminous Supersoft X-Ray Sources as Progenitors of Type Ia Supernovae

R. Di Stefano

Harvard-Smithsonian Center for Astrophysics, Cambridge, MA 02138

**Abstract.** In some luminous supersoft X-ray sources, hydrogen accretes onto the surface of a white dwarf at rates more-or-less compatible with steady nuclear burning. The white dwarfs in these systems therefore have a good chance to grow in mass. Here we review what is known about the rate of Type Ia supernovae that may be associated with SSSs. Observable consequences of the conjecture that SSSs can be progenitors of Type Ia supernovae are also discussed.

## 1 Introduction

### 1.1 The Quest for Type Ia Supernovae and Their Progenitors

Type Ia supernovae can provide important clues about the age and evolution of the Universe. Several searches expected to significantly increase the discovery rate of Type Ia supernovae are underway (see, e.g., Leibundgut et al. 1995, and Perlmutter et al. 1995). The goal of the searches is to use these bright events to measure cosmological parameters, particularly the Hubble constant, $H_0$, and the deceleration parameter, $q_0$. The success of these programs depends upon having a good understanding of the characteristics of Type Ia supernova explosions. Of particular interest is the extent to which the maximum flux, light curve profile, and spectral characteristics are uniform among Type Ia supernovae, and the ability to quantify variations. To this end, an understanding of the progenitor systems and of variations among progenitors would be important. Yet the fundamental nature of the progenitors remains mysterious. We don't even know whether the progenitors are all of one type, or whether there may be several different types of progenitor. Livio (1996) has provided us with a comprehensive review of progenitor models. Other recent reviews include those by Wheeler (1996) and Branch et al. 1995.

### 1.2 Luminous Supersoft X-Ray Sources as Type Ia Progenitors

Rappaport, Di Stefano, & Smith (RDS; 1994) proposed that close-binary supersoft sources (CBSSs) might be Type Ia progenitors. They found that reliable calculations of the rate of Type Ia supernovae that might be associated with CBSSs required a much better understanding of the evolution of the systems than was available at the time. To derive a first estimate they assumed (1) a constant accretion rate, (2) conservative mass transfer, and (3) that the total mass of the white dwarf needed to grow to $1.4 M_\odot$ for a supernova to occur. If it was further assumed that the accretion rate needed to be within the range of rates compatible with steady burning throughout the evolution, the computed rate was less than a tenth of that required. On the other hand, relaxing this condition could yield rates in the requisite range. Thus, although conclusive results were not obtained, the possibility was open that CBSSs could contribute substantially to the rate of Type Ia supernovae. Yungelson et al. (YLTTF; 1996) took a somewhat different approach, and derived supernova rates compatible with the lower limits computed by RDS, as did Canal, Ruiz-Lapuente, & Burkert 1996. Although YLTTF did follow the complete evolution of some systems, neither their calculations nor those of RDS addressed the fundamental problems that prevented a first principle evolution to be carried out for many CBSSs.

Furthermore, neither investigation treated Roche-lobe-filling systems in which the donor was very evolved at the start of mass transfer. Such systems can have mass transfer rates in or near the steady nuclear burning region. Whereas for CBSSs, rates of this magnitude are driven by the thermal time-scale readjustment of the donor, when the donor is more evolved its nuclear evolution can push the mass transfer rate into the requisite region. We will refer to Roche-lobe-filling systems in which (1) $\dot{m}$ can be within the range for steady burning of hydrogen, and (2) the donor is initially more evolved than typical in CBSSs ($m_c(0) >\sim 0.2 M_\odot$), as wide-binary supersoft sources (WBSSs). The appearance of such systems will depend on the mass transfer rate, the mass ejection rate, and on the optical depth profile. They may or may not have the observational characteristics of SSSs or of symbiotics. Whatever the observational signature, WBSSs are characterized by the state of the donor, and the fact that there is an epoch, while the donor fills its Roche lobe, during which the mass transfer rate will allow for the more-or-less steady burning of hydrogen. The systems originally proposed by Whelan and Iben (1973) as Type Ia supernova progenitors, as well as those considered by Hachisu, Kato and Nomoto (HKN; 1996) are subsets of WBSSs.

Di Stefano et al. (DNLWR; 1996) reviewed some of the uncertainties faced in computing the rate of supernovae predicted by the close-binary supersoft model, and began to study the role of mass ejection. Similar work is ongoing for the wide-binary supersoft model (Di Stefano 1996a). This paper will focus on study of the close-binary supersoft sources; the paper by Di Stefano and Nelson (DN, 1996a) serves as a companion paper which includes much of the background touched on more lightly here. Before proceeding with the details of completed and ongoing work, however, it is worth taking a moment to review the context in which the work takes place.

### 1.3 Promise and Problems

Although white dwarfs that achieve the Chandrasekhar mass, $M_C$, have long been thought to be progenitors of Type Ia supernovae, viable progenitor models have not been easy to devise. This, in spite of the fact that several varieties of accreting white dwarfs, especially cataclysmic variables (CVs) and symbiotics, have been the subject of intensive research during recent decades. In CVs, for example, the donor is typically a low mass star and, because the accretion rate is low, most or all of the mass it donates can be lost in hydrodynamic events associated with episodes of nuclear burning. The problem with symbiotics is different. Even though the donor may have enough mass to contribute in order to push the white dwarf over the Chandrasekhar limit, and even though the mass accretion rate can be compatible with steady nuclear burning, the mass transfer phase is generally too short-lived for most white dwarfs to reach $M_C$ (Kenyon et al. 1993). Recently, Yungelson et al. (1995) showed that wind-driven symbiotics, in which the donor does not fill its Roche lobe, are likely to make a negligible contribution to the rate of Type Ia supernovae if the white dwarf needs to achieve the Chandrasekhar mass in order for an explosion to occur.

Given these difficulties, it has been suggested that accreting white dwarfs may become Type Ia supernovae even if they do not reach $M_C$ (see, e.g., Woosley and Weaver 1994). The critical circumstance may instead be the ability to accrete in such a way as to form a helium mantle of $\sim 0.1 - 0.2 M_\odot$ around a C-O white dwarf. There has been good deal of study and discussion about these sub-Chandrasekhar progenitor models in recent years, but a consensus on the likelihood that they constitute a large fraction of the observed Type Ia supernovae has not yet emerged. However, even if it would become clear that reaching the Chandrasekhar mass is not an absolute requirement, this might not much change the rate of supernovae associated with some of the accreting white dwarf models. For example, Yungelson et al. (1995) found that, even if the accretion of as little as $0.1 M_\odot$ could lead to a supernova, symbiotics can account for at most 1/3 of the rate inferred from observations.

It was against this backdrop that luminous supersoft X-ray sources burst onto the scene. CBSSs seem, on the face of it to be perfect candidates for Type Ia supernova progenitors. A significant fraction of the donors are massive enough that they could donate sufficient mass to help their white dwarf companion achieve $M_C$. And the mass transfer rates can be within the range required for steady nuclear burning. Thus, the white dwarfs can genuinely increase in mass. Although the candidacy of CBSSs thus sounds promising, there are problems as well. In fact, the very features that allow the mass transfer rate to be high enough to be compatible with steady nuclear burning, the fact that the donor may be more massive and also slightly evolved, also makes the candidacy of CBSSs as Type Ia supernova progenitors somewhat problematic. This is because these same features tend to be associated with unstable mass transfer, so that many of the candidate systems risk a common envelope that would likely end the phase of steady accretion onto the white dwarf.

In this paper we will not be able to resolve the uncertainties. Instead we will attempt to clearly delineate them and the steps (both in rate computations and other tests of SSS models) that can be taken to narrow them.

## 2   Defining the Relevant Rates

It is important to clearly delineate the physical processes whose rates we would like to compute. The first hypothesis we would like to test is that the evolution of SSSs can lead to a rate of Chandrasekhar-mass explosions consistent with the rate of observed Type Ia supernovae. In this scenario, a C-O white dwarf accretes hydrogen from a companion in either a close-binary supersoft source (CBSS) or a wide-binary supersoft source (WBSS). The hydrogen burns to helium, but is likely to burn through to heavier elements before a helium mantle can develop. Thus, if the white dwarf started with an initial mass less than $\sim 1.2 M_\odot$, we are likely to witness a "classic" Chandrasekhar-mass Type Ia supernova explosion of a C-O white dwarf.

A second hypothesis we would like to test is that SSSs could lead to sub-Chandrasekhar-mass explosions. Presumably, this would require that a significant helium mantle would be able to develop, and may therefore be unlikely. Nevertheless we keep track of the numbers of systems in which the white dwarf accretes as much as $\sim 0.2 M_\odot$.

A third hypothesis, is that the explosions are actually triggered in CBSSs and WBSSs in which the binary evolution breaks down, and a common envelope ensues, leading to the merger of the white dwarf with the core of the donor. If the donor has a helium core at the time the common envelope commences, then the merger might lead to something like a sub-Chandrasekhar explosion. If, however, the donor has a C-O core (as could be the case for WBSSs), then the merger process could possibly produce a composite object with mass greater than or equal to $M_C$.

In practice, we find that events of all three types are associated with the evolution of CBSSs and WBSSs. It is the computation of the relative rate that is complicated by difficulties in computing the fraction of CBSSs and WBSSs that can survive as viable mass transfer binaries without experiencing a common envelope. It is interesting to note, however, that whatever the eventual breakdown of the relative rates, all of these types of events are predicted by the SSS models and should be observed.

## 3   Recent Work

### 3.1   Quantifying the Problems

As discussed by DN, the condition that the donor continuously fill its Roche lobe, together with the conservation of angular momentum, leads to an equation for $\dot{m}$, the mass loss rate of the donor of the following generic form.

$$\dot{m}\mathcal{D} = \mathcal{N} \quad (1)$$

$\mathcal{D}$ has a functional dependence on $\beta$, which is itself a function $\dot{m}$. Thus, equation (1) can be viewed as a non-linear equation for $\dot{m}$. There are problems with stability when $\mathcal{D}$ passes through zero and/or is negative. In general $\mathcal{D}$ can be written as $\mathcal{A} + \beta\mathcal{B}$. If $\mathcal{D}$ is negative for all $\beta > 0$, we will say that the system is in class I; systems in class I cannot be evolved using the standard formalism. A system will be said to be in Class II if there is a value of $\beta = \beta_{crit}$, such that $\mathcal{D}$ is positive only for $\beta < \beta_{crit}$; the evolution of systems in class II can be started, but will fail as the rate of mass transfer increases, if $\beta$ becomes equal to or exceeds $\beta_{crit}$. A system will be said to be in Class III if $\mathcal{D}$ is positive for all values of $\beta < 1$; systems in class III can be evolved from start to finish.

Using as input the systems that emerge as CBSS candidates from the population synthesis study of RDS, DNLWR found the following statistics. (1) Across a range of assumptions about the properties of primordial binaries and the value of $\alpha$, the common envelope ejection factor, the rate at which CBSS candidate systems are formed in a galaxy such as our own is $\sim 0.5 - 1.0$ per century. This is just $\sim 2 - 3$ times as large as the rate of Type Ia supernovae inferred from observations. The rate at which WBSS candidates are formed is more sensitive to input assumptions about $\alpha$, but can be comparable to the CBSS formation rate. (2) Across the same range of assumptions, we found that between $45-72\%$ of all CBSS systems were in class I and therefore could not be evolved. Between $10-20\%$ of all systems were in class II; their evolution crashed sometime after beginning, generally as the system approached the steady nuclear burning region. Between $17-36\%$ of all systems were in class III and could therefore be fully evolved. The story these statistics tell is somewhat more damning than may be obvious at first, since the systems in class III typically either have a mass ratio, $q = m/M$ (where $m$ is the mass of the donor and $M$ is the mass of the white dwarf), that is small (i.e., not much greater than unity), or else contain donors that are not very evolved. The associated mass transfer rates therefore tend to be small; the system does not spend much time in the steady nuclear burning region, and the white dwarf does not grow significantly. Thus, even though (and in some sense because) systems in class III can be followed, they tend not to be good candidates even for sub-Chandrasekhar Type Ia supernovae. Table 1 illustrates the range of results we obtained.

### 3.2  The Role of Mass Ejection

Table 1 illustrates two important features. First, for the population synthesis study of RDS, the majority of systems cannot be evolved using the standard formalism; they would seem to be candidates for a phase of common envelope evolution and possible mergers. Second, if the retention factor, $\beta$ can be small–i.e., if the white dwarf can eject incident material it cannot burn, then a large enough fraction of systems may survive as viable binaries, to allow CBSSs to account for a significant fraction of either Chandrasekhar-mass or sub-Chandrasekhar-mass explosions. This is the point illustrated by the last row of the table, in which an "optimistic" treatment was used: all systems for which $\mathcal{D}$ was less than zero, were artificially saved, until the system parameters changed enough to increase $\mathcal{D}$ above zero. This treatment is not realistic and was designed to give us an

Table 1. Classification of CBSS candidates by retention-factor

| Set | Case | $\beta_{crit} < 0$ Class I | $\beta_{crit} \in [0,1]$ Class II | $\beta_{crit} > 1$ Class III | $M \to 1.4 M_\odot$ | $\Delta M \geq 0.2$ |
|---|---|---|---|---|---|---|
| 1 | CON | 0.72 | 0.10 | 0.18 | **0.012** | **0.10** |
| 2 | CON | 0.67 | 0.12 | 0.21 | **0.009** | **0.02** |
| 3 | CON | 0.73 | 0.10 | 0.17 | **0.012** | **0.10** |
| 4 | CON | 0.53 | 0.15 | 0.32 | **0.015** | **0.16** |
| 5 | CON | 0.51 | 0.18 | 0.31 | **0.016** | **0.14** |
| 6 | CON | 0.45 | 0.20 | 0.36 | **0.018** | **0.18** |
| 6 | OPT | 0.45 | 0.20 | 0.36 | **0.55** | **0.81** |

Summary of the results of evolutionary calculations. "Set" refers to the data sets of CBSS candidates that emerge from each of 6 population synthesis studies we have carried out along the lines described in RDS. Note that there is relatively little variation among the results derived for different data sets. "Case" refers to the class of evolutionary "treatment" used to evolve the CBSS candidates. There are two classes of treatment, conservative (CON) and optimum (OPT). The numbers in each column represent the average fraction of systems that fall into the category indicated by the column headings. A treatment is characterized by the values of the parameters used in the evolution of the CBSS candidates. These include $a_1$ and $a_2$ (see DN), and the value of $\tilde{\xi}_{ad}$. In rows $1-6$, the average of the results for 9 separate conservative treatments is shown. In our standard conservative treatment, $\tilde{\xi}_{ad} = 4$, $a_1 = 2$, and $a_2 = 1$. Although the results for individual treatments are not shown, we note that the results among the conservative treatments are not generally dramatically different for different treatments. The exception is for $\tilde{\xi}_{ad} = 10$. This case tends to maximize the value of $\mathcal{D}$, so that all systems can be evolved; we find however, that the mass transfer rates tend to be so low that no system reaches $1.4 M_\odot$. In row 7, the results for the optimum treatment, which has been applied here only to data set 6, are shown. The evolutionary parameters are the same as those for the standard conservative treatment; when $\mathcal{D} < 0$, however, $\beta$ is chosen so as to set $\mathcal{D}$ equal to $\mathcal{D}_{min}$. Note that all systems in Class I and some in Class II are candidates for mergers.

upper limit. The fact that the upper limit so-derived is in the range of observed rates, illustrates the significant role played by mass ejection in the computation of the Type Ia supernovae rates.

It has since been discovered (HKN) that there are steady state solutions in which the white dwarf can eject the matter that it cannot burn. This is a potentially important result. Together with several other steps, it should help us to better quantify the Type Ia supernovae rate associated with SSSs. The additional needed developments include the following. (1) A population synthesis study which differs from that of RDS in including the effects of winds prior to the first common envelope phase. This has already been done by YLTTF, and by DN. The results are to move some systems from class I into class II. (2) The inclusion of radiation-driven winds. This allows us to evolve some systems in class II that would otherwise fail. We find, though, that this by itself does not lead to a significant increase in the number of systems in which the white dwarf accretes $\sim 0.2 M_\odot$ or more. (3) Implementing the full non-linear solution to Equation (1). This will allow us to better determine which systems in class II can actually be saved (Di Stefano 1996a). (4) Explicitly including a common envelope phase for those systems in which the binary evolution fails. This will allow us to better quantify the number of merger events expected. (5) Completing a full population synthesis study, including evolution, for wide-binary supersoft sources.

Work along these lines is underway and should help us to narrow the uncertainties in computations of the rate of Type Ia supernovae associated with luminous supersoft X-ray sources.

## 4 Predictions and Tests of the Model

A promising coincidence of computed and observed rates would not be conclusive evidence that the model is the unique correct Type Ia progenitor model. What would be needed in addition are testable physical predictions that go beyond rate computations. In this section we focus on two types of test. The first is the identification of individual progenitors, and the second is *post facto* study of supernovae and their remnants for "secondary characteristics" (Branch et al. 1995) that may be related to properties of the progenitor.

### 4.1 Searching for Progenitors

One way to definitively identify progenitors is to observe a system before it experiences an explosion. The problem with this approach is the events are rare. To date, no Type Ia supernova progenitors have been identified. If the rate of events is $\sim 0.3$ per century per galaxy, we would need to have detailed prior information about $\sim 30$ galaxies to have a good chance of identifying a progenitor sometime in the next decade. To test the supersoft source progenitor models we therefore ask if the distinctive signatures of SSSs would help us to identify the site of a progenitor in a distant galaxy.

**X-Ray Observations:** As part of the study of the detectability of SSSs, Di Stefano & Rappaport (1994) seeded the Magellanic Clouds and M31 with SSSs drawn from a distribution generated by using the CBSS model. Because steady nuclear burning white dwarfs of higher mass tend to be hotter and more luminous, we found that the sources most likely to be detected in M31 were those with high-mass white dwarfs. This is illustrated in Figure 1.

Note that ROSAT's study of M31 should have detected evidence of all steady nuclear burners with $M > 1.2 M_\odot$. (See Greiner J. et al., this volume: Greiner,

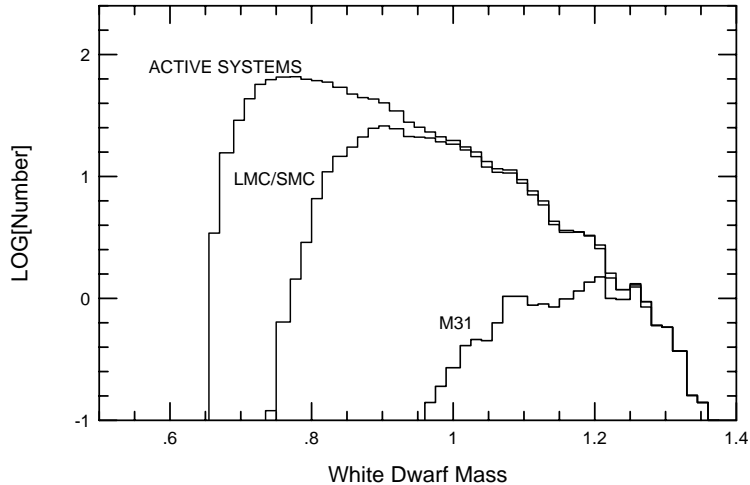

**Fig. 1.** The uppermost curve represents all active CBSSs as calculated by RDS. The middle (low) curve shows only systems that would likely have been detected by ROSAT in the LMC/SMC (M31).

Hasinger, & Kahabka 1991; Greiner, Hasinger, & Thomas 1994; Kahabka, Pietsch, Hasinger 1994; Schaeidt, Hasinger, Trümper 1993; Supper *et al* 1995; Trümper *et al.* 1991.) An important caveat is that the system should not be self-obscured, by a heavy wind, for example. This selection effect, favoring X-ray detection of systems with high-mass white dwarfs, becomes more pronounced as the distance to the host galaxy and/or absorption increases. Deep images of the most distant galaxies in which sources can be detected and resolved by X-ray satellites would therefore seem to provide potentially promising ways to identify possible progenitors. This is especially true if the Chandrasekhar-mass models are correct.

**Observations of Supersoft Nebulae**  The radiation emitted by SSSs is highly ionizing. If the sources are housed in an ISM with a local number density, $n$, of more than $\sim 1-2$ cm$^{-3}$, they may be expected to exhibit an ionization nebula with high enough surface brightness to be detected, and with distinctive properties (Rappaport, Chiang, Kallman, and Malina 1995; Chiang 1996). The central source is capable of maintaining the ionization of $\mathcal{O}(100) M_\odot$, with, for example, $\sim 2-8\%$ of the bolometric luminosity emerging in the $\lambda 5007$ line of [O III]. We will refer to these distinctive nebulae as supersoft nebulae. CAL 83 is associated with a nebula that fits the general expectations computed for a supersoft nebula (Pakull and Motch 1989; Remillard, Rappaport, and Macri 1995 [RRM]). At the detection limit of RRM, no other SSS in the Magellanic Clouds exhibits such a nebula. It is unknown what fraction (1) of the sources discovered to date, and (2) of all active SSSs, may be associated with supersoft nebulae.

DiStefano, Paerels, and Rappaport (1995; DPR) noted that at least some supersoft nebulae should have luminosities in [O III] $\lambda 5007$ comparable to the cut-off of the planetary nebula luminosity function (PNLF). (See Figure 2.)

The PNLF is used to determine extragalactic distances (see, e.g., Jacoby et al. 1992, Jacoby and Ciardullo 1993). Comparison between the SNLF and the PNLF therefore indicates that, if there are significant numbers of supersoft nebulae in distant galaxies, we should be able to detect individual SSSs in galaxies at least as far from us as the Virgo cluster. Planetary nebula surveys have been and are continuing to be carried out for dozens of galaxies. Thus, there is some chance that a coincidence between the location of a nebula and the site of a later Type Ia supernova explosion could be observed during the next decade, if SSSs can be progenitors of Type Ia supernovae (Di Stefano 1996b). Features which can help to distinguish between supersoft nebulae and planetary nebulae have been considered by DPR. It is interesting to note that supersoft nebulae are more efficient emitters in the [O III] line when the temperature of the central source is moderate, and the mass of the white dwarf is smaller than $1.2 M_\odot$. This is illustrated in Fig. 3.

Thus, while X-ray detection of SSSs in external galaxies is most likely to test and constrain Chandrasekhar-mass models, detection of supersoft nebulae in distant galaxies is most likely to test and constrain sub-Chandrasekhar-mass models.

### 4.2  Predicting Supernova Characteristics

Observational work that may allow us to eventually identify individual progenitors is exciting, but the returns are necessarily uncertain. On the other hand, we know that ongoing search programs for supernovae will certainly lead to the study of dozens of Type Ia supernovae during the next decade. Thus, if a set of tests to be applied to each observed explosion could be devised to assess the likelihood that the progenitor was a SSS, we might have a better chance of verifying or falsifying the hypothesis that SSSs are the progenitors of a significant fraction of Type Ia supernova explosions.

Branch et al. 1995 discussed a range of so-called "secondary characteristics" of supernovae that could be used to constrain progenitor models. For example, the amount and distribution of circumstellar matter can be checked via radio observations. (See, e.g., Boffi & Branch 1995, Eck et al. 1995.) Evolutionary calculations allow us to compute the total amount of mass ejected by each system and to follow the time history of mass ejection. In ongoing work, both for wide- and close-binary systems, we are therefore tracking mass ejection. Our calculations also allow us to compute the ionization state of any local ISM (as well as that of ejected material) prior to the explosion; post-explosion limits on these quantities are also possible to obtain (see, e.g., Kirshner, Winkler & Chevalier 1987, and Smith et al. 1991).

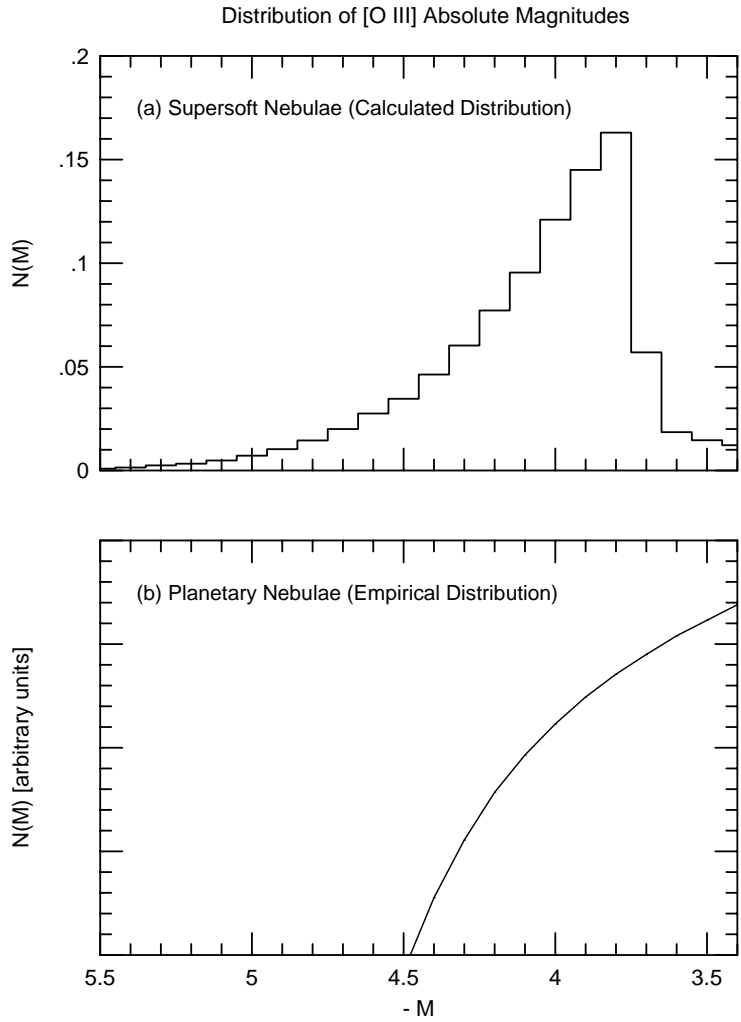

**Fig. 2.** The top panel shows the supersoft nebula luminosity function (SNLF) in [O III]. The normalization is not known; if, e.g., 10% of all SSSs have supersoft nebulae, then a total of $\mathcal{O}(100)$ supersoft nebulae may be expected to exist in a galaxy such as our own. The bottom panel shows the empirically-derived planetary nebula luminosity function (PNLF).

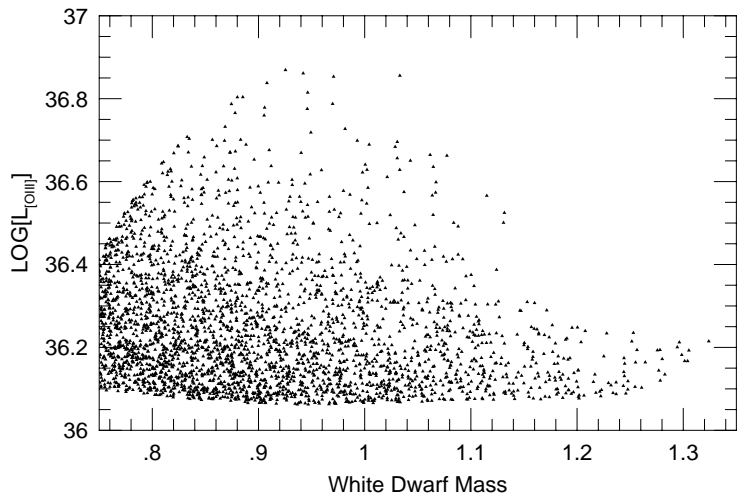

**Fig. 3.** The distribution of [O III] luminosities as a function of white dwarf mass. (Di Stefano 1996b)

## 5 Conclusions

The possibility that supersoft sources are progenitors of Type Ia supernovae is intriguing. There are many hurdles to be gotten over, however, before we can properly assess the situation.

One small hurdle has already been passed. That is, we have established that the pool of close-binary supersoft sources and wide-binary supersoft sources is large enough that, should a substantial fraction of the systems lead to supernovae, the rate of explosions could be comparable to the rate inferred from observations. This result emerges in a straightforward way from population synthesis analyses. It is interesting that the rate at which candidate progenitors are formed is, in most of our simulations, just a few times larger than the required rate. Thus, if less than 0.1 of the candidates could become supernovae, the rate of explosions due to SSSs would constitute only a small fraction of the requisite rate.

The main hurdle, then, is to determine what fraction of the candidates are actually supernova progenitors. This is a difficult problem. Solving it requires making advances in the study of the binary evolution of systems in which a more massive and possibly quite evolved star donates mass to a white dwarf companion. It seems possible that recent and ongoing work may help us to determine the fraction of candidate systems that can survive as viable binaries in which the white dwarf accretes significant mass. Even the binaries that do experience common envelopes are interesting, and determining the rates of all possible outcomes is therefore important.

Whatever the outcome of the rate calculations, the ability to evolve individual systems allows us to compute some features of the post-explosion system related to the total amount of mass ejected or to the state of ionization. Such calculations may help us to constrain the SSS models for Type Ia supernova progenitors. Further, X-ray and nebular observations of galaxies may eventually provide complementary constraints on the progenitor models.

In summary, the status of SSSs as progenitors of Type Ia supernovae is still uncertain. But there are clear lines of investigation that should help us to narrow the uncertainties.

*Acknowledgement:* I would like to thank Scott Kenyon, Lorne Nelson, Kenichi Nomoto, Saul Rappaport, and J. Craig Wheeler for interesting and useful discussions. This work has been supported in part by NSF under GER-9450087.